\documentstyle[12pt]{article}

\title{Decay on several sorts of heterogeneous centers:
Special monodisperse approximation in the situation
of strong unsymmetry. 3. Numerical results for the special monodisperse
approximation}

\author{V.Kurasov}

\date{Victor.Kurasov@pobox.spbu.ru}

\begin{document}

\maketitle

This manuscript directly continues \cite{Section1}, \cite{Section2}. All
definitions and formulas have to be taken from \cite{Section1}. The numerical
results for comparison with the total monodisperse approximation have
to be taken from \cite{Section2}.

\section{Calculations}

Now we shall turn to estimate errors of
the floating monodisperse approximation. The errors of
substitutions of the subintegral functions by the rectangular form are
known. They are rather small ($\geq 0.1$).
But the error of the floating monodisperse
approximation itself has to be estimated numerically.

Here again we can see that
the error of the number of droplets formed on the first type of heterogeneous
centers can be estimated in frame of the standard iteration method and
it is small. So, only the error in the number of the
droplets formed on the second type of heterogeneous centers will be the
subject of our interest.

Here again  the worst situation occurs when there is no
essential
exhaustion of heterogeneous centers of the second type.

We have to recall  the system of the condensation equations. Here it can
be written in  the following form
$$
G = \int_0^z \exp(-G(x)) \theta_1(x) (z-x)^3 dx
$$
$$
\theta_1 = exp(-b \int_0^z \exp(-G(x)) dx)
$$
with a positive parameter $b$ and
have to estimate the error in
$$
N = \int_0^{\infty} \exp(-l G(x)) dx
$$
with some parameter $l$.

We shall solve this problem numerically  and compare our result with
the already formulated
models. In the model of the total monodisperse approximation we get
$$
N_A = \int_0^{\infty} \exp(-l G_A(x)) dx
$$
where
$G_A$ is
$$
G_A = \frac{1}{b} (1 - \exp(-b D)) x^3
$$
and the constant $D$ is given by
$$
D = \int_0^{\infty} \exp(-x^4 /4) dx = 1.28
$$
Numerical results are shown in \cite{Section2}.

In the
model of the floating monodisperse approximation we have to calculate
the integral
$$
N_B = \int_0^{\infty} \exp(-l G_B(x)) dx
$$
where
$G_B$ is
$$
G_B = \frac{1}{b} (1 - \exp(-b \int_0^{z/4} \exp(-x^4/4) dx )) z^3
$$
$$
G_B \approx \frac{1}{b} (1 - \exp(-b (\Theta(D-z/4) z/4 + \Theta(z/4-D) D) ))
z^3
$$

We have tried all mentioned approximations for $b$ from $0.2$ up to $5.2$
with the step $0.2$ and for $l$ from $0.2$ up to $5.2$ with a step $0.2$.
We calculate the relative error in $N$. The results are drawn in fig.1
 for $N_B$  where the relative errors
are marked by  $r_2$.

The maximum of errors in $N_B$  lies near $l=0$. So, we have
to analyse the situation with small values of $l$. It was done in fig.2
  for $N_B$. We see that we can not
find the maximum error. It lies near $b=0$.
 Then we have to calculate the situation with $b=0$.
The value of $l$ can not be put directly to $l=0$.
Then  we have to solve the following equation
$$
G = \int_0^{\infty} \exp(-G(x)) (z-x)^3 dx
$$
and to compare
$$
N = \int_0^{\infty} \exp(-l G) dx
$$
with
$$
N_A = \int_0^{\infty}
\exp(-l D z^3)        dz
$$
$$
N_B = \int_0^{\infty}
\exp(-l (\Theta(z/4 -D) D z^3 + \Theta(D-z/4) z^4 /4 )  )    dz
$$

Results of this calculation will be presented together with consideration
of the "essential asymptotes" in the next section.

\pagebreak

\begin{picture}(350,350)
\put(0,0){\line(0,1){350}}
\put(0,0){\line(1,0){350}}
\put(350,0){\line(0,1){350}}
\put(0,350){\line(1,0){350}}
\put(152,206){.}
\put(158,208){.}
\put(164,210){.}
\put(170,212){.}
\put(176,214){.}
\put(182,215){.}
\put(188,217){.}
\put(194,218){.}
\put(200,219){.}
\put(206,219){.}
\put(212,220){.}
\put(218,220){.}
\put(224,221){.}
\put(230,221){.}
\put(236,221){.}
\put(242,221){.}
\put(248,220){.}
\put(254,220){.}
\put(260,220){.}
\put(266,219){.}
\put(272,219){.}
\put(278,218){.}
\put(284,218){.}
\put(290,217){.}
\put(296,216){.}
\put(302,216){.}
\put(148,199){.}
\put(154,201){.}
\put(160,203){.}
\put(166,205){.}
\put(172,207){.}
\put(178,208){.}
\put(184,210){.}
\put(190,211){.}
\put(196,212){.}
\put(202,213){.}
\put(208,214){.}
\put(214,215){.}
\put(220,215){.}
\put(226,216){.}
\put(232,216){.}
\put(238,216){.}
\put(244,216){.}
\put(250,216){.}
\put(256,216){.}
\put(262,216){.}
\put(268,216){.}
\put(274,215){.}
\put(280,215){.}
\put(286,214){.}
\put(292,214){.}
\put(298,214){.}
\put(143,194){.}
\put(149,196){.}
\put(155,198){.}
\put(161,199){.}
\put(167,201){.}
\put(173,203){.}
\put(179,204){.}
\put(185,205){.}
\put(191,207){.}
\put(197,208){.}
\put(203,208){.}
\put(209,209){.}
\put(215,210){.}
\put(221,210){.}
\put(227,211){.}
\put(233,211){.}
\put(239,211){.}
\put(245,212){.}
\put(251,212){.}
\put(257,212){.}
\put(263,211){.}
\put(269,211){.}
\put(275,211){.}
\put(281,211){.}
\put(287,211){.}
\put(293,210){.}
\put(139,189){.}
\put(145,191){.}
\put(151,193){.}
\put(157,194){.}
\put(163,196){.}
\put(169,198){.}
\put(175,199){.}
\put(181,200){.}
\put(187,201){.}
\put(193,202){.}
\put(199,203){.}
\put(205,204){.}
\put(211,205){.}
\put(217,206){.}
\put(223,206){.}
\put(229,206){.}
\put(235,207){.}
\put(241,207){.}
\put(247,207){.}
\put(253,207){.}
\put(259,207){.}
\put(265,207){.}
\put(271,207){.}
\put(277,207){.}
\put(283,207){.}
\put(289,206){.}
\put(135,184){.}
\put(141,186){.}
\put(147,188){.}
\put(153,190){.}
\put(159,191){.}
\put(165,193){.}
\put(171,194){.}
\put(177,195){.}
\put(183,197){.}
\put(189,198){.}
\put(195,199){.}
\put(201,199){.}
\put(207,200){.}
\put(213,201){.}
\put(219,201){.}
\put(225,202){.}
\put(231,202){.}
\put(237,202){.}
\put(243,203){.}
\put(249,203){.}
\put(255,203){.}
\put(261,203){.}
\put(267,203){.}
\put(273,203){.}
\put(279,203){.}
\put(285,202){.}
\put(131,180){.}
\put(137,182){.}
\put(143,183){.}
\put(149,185){.}
\put(155,187){.}
\put(161,188){.}
\put(167,189){.}
\put(173,191){.}
\put(179,192){.}
\put(185,193){.}
\put(191,194){.}
\put(197,195){.}
\put(203,195){.}
\put(209,196){.}
\put(215,197){.}
\put(221,197){.}
\put(227,198){.}
\put(233,198){.}
\put(239,198){.}
\put(245,198){.}
\put(251,199){.}
\put(257,199){.}
\put(263,199){.}
\put(269,199){.}
\put(275,198){.}
\put(281,198){.}
\put(126,175){.}
\put(132,177){.}
\put(138,179){.}
\put(144,180){.}
\put(150,182){.}
\put(156,183){.}
\put(162,185){.}
\put(168,186){.}
\put(174,187){.}
\put(180,188){.}
\put(186,189){.}
\put(192,190){.}
\put(198,191){.}
\put(204,191){.}
\put(210,192){.}
\put(216,193){.}
\put(222,193){.}
\put(228,193){.}
\put(234,194){.}
\put(240,194){.}
\put(246,194){.}
\put(252,194){.}
\put(258,194){.}
\put(264,194){.}
\put(270,194){.}
\put(276,194){.}
\put(122,171){.}
\put(128,173){.}
\put(134,174){.}
\put(140,176){.}
\put(146,177){.}
\put(152,179){.}
\put(158,180){.}
\put(164,181){.}
\put(170,182){.}
\put(176,184){.}
\put(182,184){.}
\put(188,185){.}
\put(194,186){.}
\put(200,187){.}
\put(206,187){.}
\put(212,188){.}
\put(218,189){.}
\put(224,189){.}
\put(230,189){.}
\put(236,190){.}
\put(242,190){.}
\put(248,190){.}
\put(254,190){.}
\put(260,190){.}
\put(266,190){.}
\put(272,190){.}
\put(118,167){.}
\put(124,168){.}
\put(130,170){.}
\put(136,171){.}
\put(142,173){.}
\put(148,174){.}
\put(154,176){.}
\put(160,177){.}
\put(166,178){.}
\put(172,179){.}
\put(178,180){.}
\put(184,181){.}
\put(190,182){.}
\put(196,182){.}
\put(202,183){.}
\put(208,184){.}
\put(214,184){.}
\put(220,184){.}
\put(226,185){.}
\put(232,185){.}
\put(238,185){.}
\put(244,185){.}
\put(250,186){.}
\put(256,186){.}
\put(262,186){.}
\put(268,186){.}
\put(114,162){.}
\put(120,164){.}
\put(126,166){.}
\put(132,167){.}
\put(138,168){.}
\put(144,170){.}
\put(150,171){.}
\put(156,172){.}
\put(162,173){.}
\put(168,174){.}
\put(174,175){.}
\put(180,176){.}
\put(186,177){.}
\put(192,178){.}
\put(198,178){.}
\put(204,179){.}
\put(210,180){.}
\put(216,180){.}
\put(222,180){.}
\put(228,181){.}
\put(234,181){.}
\put(240,181){.}
\put(246,181){.}
\put(252,181){.}
\put(258,181){.}
\put(264,181){.}
\put(109,158){.}
\put(115,160){.}
\put(121,161){.}
\put(127,163){.}
\put(133,164){.}
\put(139,165){.}
\put(145,167){.}
\put(151,168){.}
\put(157,169){.}
\put(163,170){.}
\put(169,171){.}
\put(175,172){.}
\put(181,173){.}
\put(187,173){.}
\put(193,174){.}
\put(199,175){.}
\put(205,175){.}
\put(211,176){.}
\put(217,176){.}
\put(223,176){.}
\put(229,177){.}
\put(235,177){.}
\put(241,177){.}
\put(247,177){.}
\put(253,177){.}
\put(259,177){.}
\put(105,154){.}
\put(111,155){.}
\put(117,157){.}
\put(123,158){.}
\put(129,160){.}
\put(135,161){.}
\put(141,162){.}
\put(147,163){.}
\put(153,164){.}
\put(159,165){.}
\put(165,166){.}
\put(171,167){.}
\put(177,168){.}
\put(183,169){.}
\put(189,170){.}
\put(195,170){.}
\put(201,171){.}
\put(207,171){.}
\put(213,172){.}
\put(219,172){.}
\put(225,172){.}
\put(231,172){.}
\put(237,173){.}
\put(243,173){.}
\put(249,173){.}
\put(255,173){.}
\put(101,149){.}
\put(107,151){.}
\put(113,152){.}
\put(119,154){.}
\put(125,155){.}
\put(131,157){.}
\put(137,158){.}
\put(143,159){.}
\put(149,160){.}
\put(155,161){.}
\put(161,162){.}
\put(167,163){.}
\put(173,164){.}
\put(179,164){.}
\put(185,165){.}
\put(191,166){.}
\put(197,166){.}
\put(203,167){.}
\put(209,167){.}
\put(215,167){.}
\put(221,168){.}
\put(227,168){.}
\put(233,168){.}
\put(239,168){.}
\put(245,168){.}
\put(251,169){.}
\put(97,145){.}
\put(103,147){.}
\put(109,148){.}
\put(115,150){.}
\put(121,151){.}
\put(127,152){.}
\put(133,153){.}
\put(139,155){.}
\put(145,156){.}
\put(151,157){.}
\put(157,158){.}
\put(163,158){.}
\put(169,159){.}
\put(175,160){.}
\put(181,161){.}
\put(187,161){.}
\put(193,162){.}
\put(199,162){.}
\put(205,163){.}
\put(211,163){.}
\put(217,163){.}
\put(223,164){.}
\put(229,164){.}
\put(235,164){.}
\put(241,164){.}
\put(247,164){.}
\put(92,141){.}
\put(98,142){.}
\put(104,144){.}
\put(110,145){.}
\put(116,147){.}
\put(122,148){.}
\put(128,149){.}
\put(134,150){.}
\put(140,151){.}
\put(146,152){.}
\put(152,153){.}
\put(158,154){.}
\put(164,155){.}
\put(170,156){.}
\put(176,156){.}
\put(182,157){.}
\put(188,157){.}
\put(194,158){.}
\put(200,158){.}
\put(206,159){.}
\put(212,159){.}
\put(218,159){.}
\put(224,160){.}
\put(230,160){.}
\put(236,160){.}
\put(242,160){.}
\put(88,137){.}
\put(94,138){.}
\put(100,139){.}
\put(106,141){.}
\put(112,142){.}
\put(118,143){.}
\put(124,145){.}
\put(130,146){.}
\put(136,147){.}
\put(142,148){.}
\put(148,149){.}
\put(154,150){.}
\put(160,150){.}
\put(166,151){.}
\put(172,152){.}
\put(178,152){.}
\put(184,153){.}
\put(190,154){.}
\put(196,154){.}
\put(202,154){.}
\put(208,155){.}
\put(214,155){.}
\put(220,155){.}
\put(226,155){.}
\put(232,156){.}
\put(238,156){.}
\put(84,132){.}
\put(90,134){.}
\put(96,135){.}
\put(102,137){.}
\put(108,138){.}
\put(114,139){.}
\put(120,140){.}
\put(126,141){.}
\put(132,142){.}
\put(138,143){.}
\put(144,144){.}
\put(150,145){.}
\put(156,146){.}
\put(162,147){.}
\put(168,147){.}
\put(174,148){.}
\put(180,149){.}
\put(186,149){.}
\put(192,150){.}
\put(198,150){.}
\put(204,150){.}
\put(210,151){.}
\put(216,151){.}
\put(222,151){.}
\put(228,151){.}
\put(234,151){.}
\put(80,128){.}
\put(86,129){.}
\put(92,131){.}
\put(98,132){.}
\put(104,134){.}
\put(110,135){.}
\put(116,136){.}
\put(122,137){.}
\put(128,138){.}
\put(134,139){.}
\put(140,140){.}
\put(146,141){.}
\put(152,142){.}
\put(158,142){.}
\put(164,143){.}
\put(170,144){.}
\put(176,144){.}
\put(182,145){.}
\put(188,145){.}
\put(194,146){.}
\put(200,146){.}
\put(206,146){.}
\put(212,147){.}
\put(218,147){.}
\put(224,147){.}
\put(230,147){.}
\put(75,124){.}
\put(81,125){.}
\put(87,127){.}
\put(93,128){.}
\put(99,129){.}
\put(105,130){.}
\put(111,132){.}
\put(117,133){.}
\put(123,134){.}
\put(129,135){.}
\put(135,136){.}
\put(141,136){.}
\put(147,137){.}
\put(153,138){.}
\put(159,139){.}
\put(165,139){.}
\put(171,140){.}
\put(177,140){.}
\put(183,141){.}
\put(189,141){.}
\put(195,142){.}
\put(201,142){.}
\put(207,142){.}
\put(213,142){.}
\put(219,143){.}
\put(225,143){.}
\put(71,120){.}
\put(77,121){.}
\put(83,122){.}
\put(89,124){.}
\put(95,125){.}
\put(101,126){.}
\put(107,127){.}
\put(113,128){.}
\put(119,129){.}
\put(125,130){.}
\put(131,131){.}
\put(137,132){.}
\put(143,133){.}
\put(149,134){.}
\put(155,134){.}
\put(161,135){.}
\put(167,136){.}
\put(173,136){.}
\put(179,136){.}
\put(185,137){.}
\put(191,137){.}
\put(197,138){.}
\put(203,138){.}
\put(209,138){.}
\put(215,138){.}
\put(221,138){.}
\put(67,115){.}
\put(73,117){.}
\put(79,118){.}
\put(85,119){.}
\put(91,121){.}
\put(97,122){.}
\put(103,123){.}
\put(109,124){.}
\put(115,125){.}
\put(121,126){.}
\put(127,127){.}
\put(133,128){.}
\put(139,129){.}
\put(145,129){.}
\put(151,130){.}
\put(157,131){.}
\put(163,131){.}
\put(169,132){.}
\put(175,132){.}
\put(181,133){.}
\put(187,133){.}
\put(193,133){.}
\put(199,134){.}
\put(205,134){.}
\put(211,134){.}
\put(217,134){.}
\put(63,111){.}
\put(69,112){.}
\put(75,114){.}
\put(81,115){.}
\put(87,116){.}
\put(93,117){.}
\put(99,119){.}
\put(105,120){.}
\put(111,121){.}
\put(117,122){.}
\put(123,123){.}
\put(129,123){.}
\put(135,124){.}
\put(141,125){.}
\put(147,126){.}
\put(153,126){.}
\put(159,127){.}
\put(165,127){.}
\put(171,128){.}
\put(177,128){.}
\put(183,129){.}
\put(189,129){.}
\put(195,129){.}
\put(201,129){.}
\put(207,130){.}
\put(213,130){.}
\put(58,107){.}
\put(64,108){.}
\put(70,109){.}
\put(76,111){.}
\put(82,112){.}
\put(88,113){.}
\put(94,114){.}
\put(100,115){.}
\put(106,116){.}
\put(112,117){.}
\put(118,118){.}
\put(124,119){.}
\put(130,120){.}
\put(136,121){.}
\put(142,121){.}
\put(148,122){.}
\put(154,122){.}
\put(160,123){.}
\put(166,123){.}
\put(172,124){.}
\put(178,124){.}
\put(184,125){.}
\put(190,125){.}
\put(196,125){.}
\put(202,125){.}
\put(208,126){.}
\put(54,103){.}
\put(60,104){.}
\put(66,105){.}
\put(72,106){.}
\put(78,108){.}
\put(84,109){.}
\put(90,110){.}
\put(96,111){.}
\put(102,112){.}
\put(108,113){.}
\put(114,114){.}
\put(120,115){.}
\put(126,115){.}
\put(132,116){.}
\put(138,117){.}
\put(144,118){.}
\put(150,118){.}
\put(156,119){.}
\put(162,119){.}
\put(168,120){.}
\put(174,120){.}
\put(180,120){.}
\put(186,121){.}
\put(192,121){.}
\put(198,121){.}
\put(204,121){.}
\put(50,98){.}
\put(56,100){.}
\put(62,101){.}
\put(68,102){.}
\put(74,103){.}
\put(80,105){.}
\put(86,106){.}
\put(92,107){.}
\put(98,108){.}
\put(104,109){.}
\put(110,109){.}
\put(116,110){.}
\put(122,111){.}
\put(128,112){.}
\put(134,113){.}
\put(140,113){.}
\put(146,114){.}
\put(152,114){.}
\put(158,115){.}
\put(164,115){.}
\put(170,116){.}
\put(176,116){.}
\put(182,116){.}
\put(188,117){.}
\put(194,117){.}
\put(200,117){.}
\put(46,94){.}
\put(52,95){.}
\put(58,97){.}
\put(64,98){.}
\put(70,99){.}
\put(76,100){.}
\put(82,101){.}
\put(88,102){.}
\put(94,103){.}
\put(100,104){.}
\put(106,105){.}
\put(112,106){.}
\put(118,107){.}
\put(124,108){.}
\put(130,108){.}
\put(136,109){.}
\put(142,109){.}
\put(148,110){.}
\put(154,110){.}
\put(160,111){.}
\put(166,111){.}
\put(172,112){.}
\put(178,112){.}
\put(184,112){.}
\put(190,112){.}
\put(196,113){.}
\put(150,200){\vector(0,1){100}}
\put(150,200){\vector(1,0){170}}
\put(150,200){\vector(-1,-1){120}}
\put(140,300){$r_2$}
\put(335,205){$b$}
\put(20,70){$l$}
\put(40,90){\line(1,0){156}}
\put(40,90){\line(0,1){4}}
\put(20,90){5.2}
\put(306,200){\line(-1,-1){110}}
\put(306,200){\line(0,1){20}}
\put(306,215){5.2}
\put(196,90){\line(0,1){23}}
\put(125,245){0.1}
\end{picture}

{ \it

\begin{center}
Fig.1
\end{center}

 The relative error of $N_B$ drawn as the function of $l$ and $b$.
Parameter $l$ goes from $0.2$ up to $5.2$ with a step $0.2$.
Parameter $b$ goes from $0.2$ up to $5.2$ with a step $0.2$.

One can see the maximum at small $l$ and moderate $b$.

}

\pagebreak

\begin{picture}(350,350)
\put(0,0){\line(0,1){350}}
\put(0,0){\line(1,0){350}}
\put(350,0){\line(0,1){350}}
\put(0,350){\line(1,0){350}}
\put(145,229){.}
\put(151,229){.}
\put(157,228){.}
\put(163,227){.}
\put(169,226){.}
\put(175,224){.}
\put(181,223){.}
\put(187,221){.}
\put(193,219){.}
\put(199,217){.}
\put(205,215){.}
\put(211,213){.}
\put(217,211){.}
\put(223,210){.}
\put(229,208){.}
\put(235,207){.}
\put(241,206){.}
\put(247,204){.}
\put(253,203){.}
\put(259,202){.}
\put(265,201){.}
\put(271,200){.}
\put(277,199){.}
\put(283,199){.}
\put(289,198){.}
\put(295,197){.}
\put(135,209){.}
\put(141,210){.}
\put(147,211){.}
\put(153,211){.}
\put(159,211){.}
\put(165,210){.}
\put(171,210){.}
\put(177,209){.}
\put(183,208){.}
\put(189,207){.}
\put(195,205){.}
\put(201,204){.}
\put(207,203){.}
\put(213,201){.}
\put(219,200){.}
\put(225,199){.}
\put(231,198){.}
\put(237,197){.}
\put(243,195){.}
\put(249,194){.}
\put(255,193){.}
\put(261,192){.}
\put(267,192){.}
\put(273,191){.}
\put(279,190){.}
\put(285,189){.}
\put(124,194){.}
\put(130,195){.}
\put(136,196){.}
\put(142,197){.}
\put(148,197){.}
\put(154,197){.}
\put(160,197){.}
\put(166,197){.}
\put(172,196){.}
\put(178,196){.}
\put(184,195){.}
\put(190,194){.}
\put(196,193){.}
\put(202,192){.}
\put(208,191){.}
\put(214,190){.}
\put(220,189){.}
\put(226,188){.}
\put(232,187){.}
\put(238,186){.}
\put(244,185){.}
\put(250,184){.}
\put(256,183){.}
\put(262,182){.}
\put(268,181){.}
\put(274,180){.}
\put(114,180){.}
\put(120,182){.}
\put(126,183){.}
\put(132,184){.}
\put(138,185){.}
\put(144,185){.}
\put(150,186){.}
\put(156,186){.}
\put(162,185){.}
\put(168,185){.}
\put(174,184){.}
\put(180,184){.}
\put(186,183){.}
\put(192,182){.}
\put(198,181){.}
\put(204,180){.}
\put(210,179){.}
\put(216,178){.}
\put(222,177){.}
\put(228,176){.}
\put(234,175){.}
\put(240,174){.}
\put(246,173){.}
\put(252,172){.}
\put(258,172){.}
\put(264,171){.}
\put(103,167){.}
\put(109,169){.}
\put(115,171){.}
\put(121,172){.}
\put(127,173){.}
\put(133,173){.}
\put(139,174){.}
\put(145,174){.}
\put(151,174){.}
\put(157,174){.}
\put(163,174){.}
\put(169,173){.}
\put(175,172){.}
\put(181,172){.}
\put(187,171){.}
\put(193,170){.}
\put(199,169){.}
\put(205,168){.}
\put(211,167){.}
\put(217,166){.}
\put(223,165){.}
\put(229,165){.}
\put(235,164){.}
\put(241,163){.}
\put(247,162){.}
\put(253,161){.}
\put(92,155){.}
\put(98,157){.}
\put(104,158){.}
\put(110,160){.}
\put(116,161){.}
\put(122,162){.}
\put(128,163){.}
\put(134,163){.}
\put(140,163){.}
\put(146,163){.}
\put(152,163){.}
\put(158,162){.}
\put(164,162){.}
\put(170,161){.}
\put(176,161){.}
\put(182,160){.}
\put(188,159){.}
\put(194,158){.}
\put(200,157){.}
\put(206,156){.}
\put(212,156){.}
\put(218,155){.}
\put(224,154){.}
\put(230,153){.}
\put(236,152){.}
\put(242,151){.}
\put(82,143){.}
\put(88,145){.}
\put(94,147){.}
\put(100,148){.}
\put(106,149){.}
\put(112,150){.}
\put(118,151){.}
\put(124,152){.}
\put(130,152){.}
\put(136,152){.}
\put(142,152){.}
\put(148,152){.}
\put(154,151){.}
\put(160,151){.}
\put(166,150){.}
\put(172,150){.}
\put(178,149){.}
\put(184,148){.}
\put(190,147){.}
\put(196,146){.}
\put(202,146){.}
\put(208,145){.}
\put(214,144){.}
\put(220,143){.}
\put(226,142){.}
\put(232,141){.}
\put(71,131){.}
\put(77,133){.}
\put(83,135){.}
\put(89,137){.}
\put(95,138){.}
\put(101,139){.}
\put(107,140){.}
\put(113,141){.}
\put(119,141){.}
\put(125,141){.}
\put(131,141){.}
\put(137,141){.}
\put(143,141){.}
\put(149,140){.}
\put(155,140){.}
\put(161,139){.}
\put(167,139){.}
\put(173,138){.}
\put(179,137){.}
\put(185,136){.}
\put(191,135){.}
\put(197,135){.}
\put(203,134){.}
\put(209,133){.}
\put(215,132){.}
\put(221,131){.}
\put(61,120){.}
\put(67,122){.}
\put(73,124){.}
\put(79,125){.}
\put(85,127){.}
\put(91,128){.}
\put(97,129){.}
\put(103,130){.}
\put(109,130){.}
\put(115,130){.}
\put(121,130){.}
\put(127,130){.}
\put(133,130){.}
\put(139,130){.}
\put(145,129){.}
\put(151,129){.}
\put(157,128){.}
\put(163,127){.}
\put(169,127){.}
\put(175,126){.}
\put(181,125){.}
\put(187,124){.}
\put(193,124){.}
\put(199,123){.}
\put(205,122){.}
\put(211,121){.}
\put(50,108){.}
\put(56,111){.}
\put(62,112){.}
\put(68,114){.}
\put(74,116){.}
\put(80,117){.}
\put(86,118){.}
\put(92,119){.}
\put(98,119){.}
\put(104,119){.}
\put(110,120){.}
\put(116,120){.}
\put(122,119){.}
\put(128,119){.}
\put(134,119){.}
\put(140,118){.}
\put(146,118){.}
\put(152,117){.}
\put(158,116){.}
\put(164,116){.}
\put(170,115){.}
\put(176,114){.}
\put(182,113){.}
\put(188,113){.}
\put(194,112){.}
\put(200,111){.}
\put(39,97){.}
\put(45,99){.}
\put(51,101){.}
\put(57,103){.}
\put(63,104){.}
\put(69,106){.}
\put(75,107){.}
\put(81,108){.}
\put(87,108){.}
\put(93,109){.}
\put(99,109){.}
\put(105,109){.}
\put(111,109){.}
\put(117,109){.}
\put(123,108){.}
\put(129,108){.}
\put(135,107){.}
\put(141,107){.}
\put(147,106){.}
\put(153,105){.}
\put(159,105){.}
\put(165,104){.}
\put(171,103){.}
\put(177,102){.}
\put(183,102){.}
\put(189,101){.}
\put(150,200){\vector(0,1){100}}
\put(150,200){\vector(1,0){170}}
\put(150,200){\vector(-1,-1){120}}
\put(140,300){$r_3$}
\put(335,205){$b$}
\put(20,70){$l$}
\put(33,83){\line(1,0){156}}
\put(33,83){\line(0,1){14}}
\put(13,83){0.1}
\put(306,200){\line(-1,-1){117}}
\put(306,200){\line(0,1){8}}
\put(311,215){5.2}
\put(189,83){\line(0,1){18}}
\put(125,245){0.1}
\end{picture}

\begin{center}
Fig.2
\end{center}

{ \it
 The relative error of $N_B$ drawn as the function of $l$ and $b$.
Parameter $l$ goes from $0.01$ up to $0.11$ with a step $0.01$.
Parameter $b$ goes from $0.2$ up to $5.2$ with a step $0.2$.

One can see the maximum at small $l$ and small $b$. One can note that
now the values of $b$ corresponding to maximum of the relative errors
become small.

}


\begin{thebibliography}{99}

\bibitem{Section1}
V. Kurasov, Preprint cond-mat@xxx.lanl.gov get 0001104

\bibitem{Section2}
V. Kurasov, Preprint cond-mat@xxx.lanl.gov get 0001108


\end{thebibliography}
\end{document}